\documentclass[prd, reprint, amsmath, amssymb, floatfix, aps, longbibliography, superscriptaddress]{revtex4-2}
\usepackage{graphicx}
\usepackage{dcolumn}
\usepackage{bm}
\usepackage[colorlinks=true,urlcolor=rossos,anchorcolor=black,citecolor=mygreen,filecolor=black,linkcolor=black,menucolor=blue,linktocpage=true]{hyperref} 
\usepackage{subfigure}

\usepackage{booktabs}
\usepackage{colortbl}
\usepackage{collcell}
\usepackage{enumerate}
\usepackage{float}
\usepackage{verbatim}
\usepackage{multirow}
\usepackage{xparse}
\usepackage{tabularx}
\usepackage{cancel}

\graphicspath{{figs/}}

\usepackage{multirow}

\definecolor{rossos}{cmyk}{0,1,1,0.55}
\definecolor{mygreen}{rgb}{0.27, 0.64, 0.48}
\definecolor{mygray}{gray}{0.95}

\begin{document}

\title{Model of the quintessence axion}

\author{Sudhakantha Girmohanta}
\email{sgirmohanta@sjtu.edu.cn}
\affiliation{Tsung-Dao Lee Institute and School of Physics and Astronomy, \\
Shanghai Jiao Tong University, 520 Shengrong Road, Shanghai, 201210, China}

\author{Yu-Cheng Qiu}
\email{ethanqiu@sjtu.edu.cn}
\affiliation{Tsung-Dao Lee Institute and School of Physics and Astronomy, \\
Shanghai Jiao Tong University, 520 Shengrong Road, Shanghai, 201210, China}

\author{Jin-Wei Wang}
\email{jinwei.wang@uestc.edu.cn}
\affiliation{School of Physics, University of Electronic Science and Technology of China, Chengdu 611731, China}
\affiliation{Tsung-Dao Lee Institute and School of Physics and Astronomy, \\
Shanghai Jiao Tong University, 520 Shengrong Road, Shanghai, 201210, China}

\author{Tsutomu T. Yanagida}
\email{tsutomu.tyanagida@sjtu.edu.cn}
\affiliation{Tsung-Dao Lee Institute and School of Physics and Astronomy, \\
Shanghai Jiao Tong University, 520 Shengrong Road, Shanghai, 201210, China}
\affiliation{Kavli IPMU (WPI), The University of Tokyo, Kashiwa, Chiba 277-8583, Japan}

\begin{abstract}
We construct a model of the quintessence axion based on a gauged chiral $U(1)$ symmetry and an additional flat fifth dimension. The required high qualities are guaranteed by the brane separation. The observed cosmological constant (i.e., the potential energy of the quintessence axion) is determined by the size of the extra dimension and the axion decay constant $F_a$ is fixed almost at $F_a\simeq10^{17}\,{\rm GeV}$, which is sufficiently large for the stability of the axion field near the hilltop of its potential. Furthermore, the movement of the axion can also easily explain the recently reported isotropic cosmic birefringence of the cosmic microwave background photon.

\end{abstract}

\maketitle

\section{Introduction}

The space-time metric $g(x)_{\mu\nu}$ might be dynamically generated and it seems very natural that the Universe is asymptotically flat and the cosmological constant is going to vanish in the infinite future \cite{Aoki:1998vn,Dvali:2018fqu}. Therefore, it is very interesting to consider the observed cosmological constant (CC) $\Lambda\simeq (2.26\times 10^{-3}\,{\rm eV})^4$~\cite{Planck:2018vyg} as the temporal potential energy of some light-scalar boson fields.
Under this circumstance, the mass of the scalar boson must be extremely small, i.e. $\sim 10^{-33}\,{\rm eV}$, to not roll down to the potential minimum until the present. Hence, it is very natural to consider it as a Nambu-Goldstone boson~\cite{Nambu:1960tm,Goldstone:1961eq,Goldstone:1962es} (we call it the quintessence axion~\cite{Fukugita:1994hq, Nomura:2000yk}), and thus its mass is protected by a global symmetry and against the radiative corrections. However, it is believed that any global symmetries must be broken by non-perturbation effects in quantum gravity~\cite{Banks:2010zn}. We need an extremely good mechanism to protect this required global symmetry against the nonperturbative effects in quantum gravity, that is, the quintessence axion needs to be of extremely high quality. Another issue is that we do not know the origin of the quintessence axion in the UV theory.

In a recent paper~\cite{Qiu:2023los} we proposed a general framework, based on gauged chiral $U(1)$ and $\mathbf{Z}_{2N}$ symmetries~\cite{Fukuda:2017ylt,Nakayama:2011dj}, to answer both above questions. 
However, to get sufficient suppression for the dangerous high-order operator, many pairs of chiral fermions are introduced, which render a relatively small effective axion decay constant, i.e. $F_a \sim 3\times 10^{16}\,{\rm GeV}$, and make the stability of quintessence axion potential disconcerting. 
As mentioned in Ref.~\cite{Qiu:2023los}, such a $F_a$ seems to be too small to guarantee axion stability. In this short paper, we construct a consistent model of the quintessence axion by embedding our framework into a five-dimensional theory. As we shall see, compared to Ref.~\cite{Qiu:2023los}, the particle spectrum becomes much simpler and the additional gauge symmetry $\mathbf{Z}_{2N}$ is also not necessary thanks to the brane separation configuration. Note that in the following content, we have assumed a vanishing four-dimensional CC in the effective action.

The geometric structure of this model involves a $S^1/\mathbf{Z}_2$ orbifold topology for the compactification of the extra dimension and two $3$-branes that are placed on fixed points. The metric of extra dimension could be flat or warped. Moreover, we find that the flat setup is more attractive because it can produce appropriate high-quality quintessence axion~\cite{Izawa:2002qk}, while for the warped case, the suppression is too small to provide a satisfying quintessence axion candidate, but surprisingly it can be used to produce an ideal fuzzy dark matter (DM) axion (see Sec.~\ref{sec:discussion}). The brane setup is stabilized through the Goldberger-Wise mechanism~\cite{Goldberger:1999uk,Chacko:2002sb}. 

As for the particle content, adopting the same strategy in Ref.~\cite{Qiu:2023los}, we introduce two Higgs fields $\phi_i$ and
two pairs of chiral electrons $\{\psi_i,\overline{\psi}_i\}$ with $i=1,2$ and put them on two different branes~\cite{Izawa:2002qk}. As shown in Ref.~\cite{Fukuda:2017ylt}, two Higgs fields indicate two global $U(1)$ symmetries, and one of their linear combinations which is anomaly free can be gauged, called $U(1)_g$, while the other orthogonal linear combination is $U(1)_a$ that is the origin of the axion. By adopting the proper charge assignments the unwanted vectorlike mass terms can be avoided. All chiral electrons acquire mass only through the Yukawa couplings~\cite{Qiu:2023los}. Furthermore, we introduce one bulk scalar $\Phi$, which carries $U(1)_g$ charge and couples to both brane Higgs fields.
After integrating the heavy bulk scalar $\Phi$ and the extra fifth dimension, one ends up with a four-dimensional theory, where two Higgs $\phi_{1,2}$ can couple to each other with an exponentially suppressed factor, which is similar to the Yukawa potential. The suppression factor is determined by the fundamental scale of the five-dimensional theory.

At the lower energy scale, the $U(1)_g$ and $U(1)_a$ are spontaneously broken by vacuum expectation values (VEVs) of brane Higgs fields. One Nambu-Goldstone particle is absorbed by the gauge field and becomes its longitudinal mode, and the other one is the axion. Here the axion potential is generated through the exponentially suppressed coupling between two brane Higgs fields. The hilltop of axion potential can explain the observed CC. Meanwhile, the axion decay constant $F_a$ is directly linked with the fundamental scale of the extra dimension, which is large enough to ensure stability. Another interesting fact is that the quintessence axion can couple to the photon field through the Chern-Simons type term, which could explain the reported isotropic cosmic birefringence (ICB) simultaneously~\cite{Minami:2020odp}. 


\section{The Setup}
\label{sec:framework_section}

We consider a $\mathbf{R}_4\times S^1/\mathbf{Z}_2$ topology, where the extra dimension is flat. The metric is given by
\begin{equation}
    ds^2 = \eta_{\mu\nu} dx^\mu dx^\nu - dy^2\;,
    \label{metric:eq}
\end{equation}
where $\mu=0,\cdots,3$ is the $4D$ indices. Here $\eta_{\mu\nu} = {\rm diag}{(1,-1,-1,-1)}$ is the Minkowski metric. The $y$-dim is compactified and let $0\leq y \leq L$. The extra-dimensional setup entails a cutoff denoted as $M_*$.
Two $3$-branes are put on two endpoints in the $y$-dim, namely $y=0$ and $y=L$. Two pairs of chiral electrons $\{\psi_1,\overline{\psi}_1\}$ and $\{\psi_2,\overline{\psi}_2\}$ are put on two different branes separately. Here we assume that $\psi_i\in(1,1,C_{Y})$ and $\overline{\psi}_i\in(1,1,-C_{Y})$ under the Standard Model $SU(3)_{\rm c} \times SU(2)_{\rm L} \times U(1)_{\rm Y}$ gauge transformations. Besides, they are also charged under the gauged $U(1)_g$ symmetry.
Two brane Higgs fields $\phi_i$ are introduced so that the electrons can receive mass through the Yukawa couplings. Moreover, because of the $U(1)_g$ gauge invariance, the $U(1)_g$ charge of $\phi_i$ can be fixed. 
In addition, we introduce a bulk scalar $\Phi(x^\mu,y)$, which also carries the $U(1)_g$ charge and can interact with both $\phi_1$ and $\phi_2$ on two separated branes. The $U(1)_g$ charge assignments of all particles are given in Table.~\ref{tab:charges}. 

\begin{table}[t]
\caption{The $U(1)_g$ charge assignment.}
\label{tab:charges}
\begin{ruledtabular}
\begin{tabular}{ccccc}
$i$ & $\psi_i(x^\mu)$ & $\overline{\psi}_i(x^\mu)$ & $\phi_i(x^\mu)$ & $\Phi(x^\mu,y)$  \\
\midrule
$1$ & $1$ & $1$ & $-2$ &  \multirow{2}{*}{$2$} \\ [0.4em]
$2$ & $-1$ & $-1$ & $2$ &  \\
\end{tabular}
\end{ruledtabular}
\end{table}

After adopting the proper normalization such that all scalar fields have mass dimensions $1$, the $5D$ action can be expressed as
\begin{equation}
    S_5 =\int d^4 x d y \sqrt{-g} \left[ \mathcal{L}_{\rm bulk} + \mathcal{L}_1 \delta(y) + \mathcal{L}_2 \delta(y-L) \right] \;,
    \label{eq:action}
\end{equation}
where
\begin{subequations}
\begin{align}
    \mathcal{L}_{\rm bulk} & =  M_*^3 R_5 + M_*|D_a\Phi|^2 - M_*^3 |\Phi|^2 + \cdots,  \\ 
    \mathcal{L}_1 & = |D_\mu \phi_1|^2 - c_1 M_*^2  \phi_1 \Phi + \cdots, \\
    \mathcal{L}_2 & = |D_\mu \phi_2|^2 - c_2 M_*^2  \phi_2 \Phi^* +\cdots.
\end{align}
\end{subequations}
The $D_a$ is the covariant derivative that contains the gauge fields, $R_5$ is the $5D$ Ricci scalar, $c_{1,2}$ is the dimensionless order-one coupling constant. The dots contain terms of gauge field and fermions. Since we are only interested in the interactions between scalars, they are neglected for simplicity. 

There is only one fundamental scale in our $5D$ setup, and that is $M_*$. The higher dimensional gravitational action is given by $S_{\rm HE,5} = \int d^5x M_*^3 \sqrt{-g} R_5$. Note that for the flat extra dimension, with the metric given in Eq.~(\ref{metric:eq}), $R_5 = R$, where $R$ is the $4D$ Ricci scalar. By integrating over the $y$-dim and matching with the usual $4D$ Hilbert-Einstein action, i.e.  $S_{\rm HE} = \int d^4 x \sqrt{-g_4} (M_*^3 L) R =  \int d^4 x \sqrt{-g_4} M_{\rm Pl}^2 R$ with $M_{\rm Pl} = 2.4 \times 10^{18}\,{\rm GeV}$ identified as the reduced Planck scale, one can obtain that 
\begin{equation}
    M_* L = \left( \frac{M_{\rm Pl}}{M_*} \right)^2\;.
\end{equation}

Using the method of variation, the equation of motion of $\Phi$ can be obtained, that is
\begin{align}
    & \left( \eta^{\mu\nu}\partial_\mu\partial_\nu - \partial_y^2 + M_*^2 \right)\Phi(x^\mu,y) =  \nonumber\\
    &\qquad -M_*  [c_1  \phi_1^*(x^\mu)\delta(y) +  c_2  \phi_2(x^\mu)\delta(y-L)]+ \cdots \;.
    \label{eq:EOM_Phi}
\end{align}
Here the dots represent the insignificant gauge terms that do not violate $U(1)_a$, so they will not contribute to the $U(1)_a$-breaking operator.
Combining Eqs.~\eqref{eq:action} and \eqref{eq:EOM_Phi}, we can derive the low-energy effective $4D$ Lagrangian by integrating the extra dimension (see Appendix \ref{app:A} for more detailed derivations), which can be expressed as 
\begin{align}
     \mathcal{L}_{\rm eff, int} &= c_1 c_2 M_*^2 e^{-M_* L}\phi_1 (x^\mu) \phi_2 (x^\mu) + {\rm H.c.} \nonumber \\
     &= c_1 c_2 M_*^2  \phi_1  \phi_2  e^{-(M_{\rm Pl}/M_*)^2} + {\rm H.c.}\;.
    \label{eq:Leff}
\end{align}
It shows that the interaction term between $\phi_1$ and $\phi_2$ violates $U(1)_a$ and also gets suppressed by an exponential factor.
The origin of this exponential suppression factor is from the Fourier space decomposition of $\delta(y) \delta(y-L)$, which gives rise to a factor proportional to $e^{ipL}$, where $p$ is the Fourier space momentum variable. After integrating over the momentum space, this picks up the residue at $p = iM_*$, resulting in this factor of $e^{-M_* L}$ \cite{Izawa:2002qk}. This can be also thought of as the suppression due to a Yukawa-like propagator of a mediator particle whose mass lies at $M_*$.

\section{The Quintessence Axion}
\label{sec:quintessence}
In our model, the possible lowest-order operator that obeys the gauge $U(1)_g$ symmetry but breaks the global $U(1)_a$ symmetry is
\begin{equation}
\mathcal{O} = M_*^2\phi_1 \phi_2 e^{-(M_{\rm Pl}/M_*)^2} + \text{H.c.}\;.
\label{eq:PQbreak}
\end{equation}
Note that the order one coupling constant has been neglected.
After spontaneous symmetry breaking, one can expand two Higgs fields as $\phi_1 = (f_1/\sqrt{2}) \exp{(i\tilde{a}/f_1)}$ and $\phi_2 = (f_2/\sqrt{2}) \exp{(i \tilde{b}/f_2)}$, where $f_i$ is the VEV of $\phi_i$.
The axion $a$ is a linear combination of $\tilde{a}$ and $\tilde{b}$~\cite{Fukuda:2017ylt}. This operator~\eqref{eq:PQbreak} generates the potential of $a$, which is
\begin{equation}
V = \frac{\Lambda_a}{2} \left( 1- \cos{\frac{a}{F_a}}\right)\;,
\label{eq:V}
\end{equation}
where
\begin{equation}
    \Lambda_a = 2 f_1 f_2 M_*^{2} e^{-(M_{\rm Pl}/M_*)^2},\quad F_a = \frac{f_1 f_2}{\sqrt{f_1^2 + f_2^2}}. 
    \label{eq:lambdaaxion}
\end{equation}
The $\Lambda_a$ represents the potential energy at the hilltop, which can be used to explain the observed CC, and $F_a$ is the axion decay constant. Note that here we already assumed that the true vacuum has zero potential energy.

To quantitatively discuss the quality of quintessence axion, here we take $f_2 = f_1 = M_*$ as a benchmark, and the numerical evaluation gives us
\begin{equation}
    \Lambda_a \approx \Lambda \left(\frac{M_*}{1.47\times 10^{17}\,{\rm GeV}} \right)^{515}\;.
    \label{eq:lambdaa}
\end{equation}
As one could see, $\Lambda_a$ is extremely sensitive to the value of $M_*$. Therefore, to satisfy the observation, the size of the extra dimension (equally the value of $M_*$) is almost fixed. Besides, the axion decay constant is
\begin{equation}
    F_a = \frac{M_*}{\sqrt{2}} = 1.04 \times 10^{17}\,{\rm GeV},
    \label{eq:Fa}
\end{equation}
which is large enough to ensure the stability.

Since the chiral electrons $\psi_i$ and $\overline{\psi}_j$ carry $U(1)_{\rm Y}$ charges, we can show that the $[U(1)_a]\times [U(1)_{\rm Y}]^2$ anomaly is nonzero (more detailed derivations can be found in Ref.~\cite{Qiu:2023los}). After doing the anomaly matching, the Chern-Simons type term appears in the form of 
\begin{equation}
  \mathcal{L} \supset c_\gamma \frac{a}{F_a} \frac{g^2 }{16\pi^2} F_{\mu\nu}\tilde{F}^{\mu\nu}\;,
  \label{eq:photon_coupling}
\end{equation}
where $c_\gamma=C_{Y}^2$ is an anomaly coefficient, $F_{\mu\nu}$ and $\tilde{F}^{\mu\nu}$ are photon field strength and its dual. 
As shown in Ref.~\cite{Lin:2022niw,Choi:2021aze}, this quintessence axion could explain the ICB.
In order to explain the observed nonvanishing rotation angle $\beta=0.35\pm0.14$ deg~\cite{Minami:2020odp}, we should have $c_\gamma \gtrsim 15$~\cite{Lin:2022niw}. Therefore, we can choose that $C_{Y}\gtrsim4$. In addition to choosing a relatively larger hypercharge, one can also adopt the same strategy in Ref.~\cite{Qiu:2023los}, that is introducing $N$ copies pairs of electrons on each brane, which implies $c_\gamma=N C_{Y}^2$. For $C_{Y}=1$, we need $N\gtrsim 15$ to explain the ICB.
This $N$ is similar to the concept of family number in the Standard Model. Note that we have checked that neither of these strategies will produce a Landau pole of $U(1)_{Y}$ gauge coupling.


\section{Summary and Discussion}
\label{sec:discussion}
As an extension of our previous work~\cite{Qiu:2023los}, in this paper, we propose a new quintessence axion model by introducing a chiral $U(1)$ gauge symmetry and the flat fifth extra dimension. More specifically, the new Higgs fields $\phi_i$ and fermion pairs $\{\psi_i,\overline{\psi}_i\}$ are placed on two separate branes. Integrating out the charged bulk field $\Phi$, we find that the lowest Peccei-Quinn (PQ) breaking operator gets suppressed by a factor of $\sim e^{-(M_{\rm Pl}/M_*)^2}$, which can provide us an appropriate quintessence axion. 
A key observation is that the decay constant $F_a$ is now linked with the brane separation scale, i.e. $M_*$.
Compared with Ref.~\cite{Qiu:2023los}, our current model has several advantages: (1) the gauged $\mathbf{Z}_{2N}$ is not needed because of the brane separation setup; (2) the suppression factor is independent of pairs of fermions, so not too many fermions are needed; (3) the $F_a$ can be as large as $\sim 10^{17}\,{\rm GeV}$, so there is no instability problem; (4) the $U(1)_g$ charge assignment is simpler.

Moreover, our model can also easily explain the observed ICB by setting $C_{Y}\gtrsim4$ or introducing $N\gtrsim 15$ copies of the fermion pairs on each brane. One important merit of this model is that the axion potential $V$ and axion decay constant $F_a$ are independent of $N$, thus more pairs of fermion will not affect the results of Eqs.~\eqref{eq:lambdaa}--\eqref{eq:Fa}.

Here the suppression factor comes from the brane separation and the effective interaction between $\phi_1$ and $\phi_2$ is generated from bulk mediator $\Phi$. One could also consider a special geometry like a wormhole to connect two branes and thus generate interaction like~\eqref{eq:PQbreak}, which is also exponentially suppressed by the wormhole action~\cite{Kallosh:1995hi}. In this case, the bulk $\Phi$ is absent and the goal could also be achieved.

Another interesting fact is that in this work we only discuss the flat extra-dimension case, while for the warped case, the suppression factor becomes $\sim e^{-M_{\rm Pl}/M_*}$, where $M_*$ is now the mass scale associated with the infrared (IR) brane~\cite{Goldberger:2002cz,Fichet:2019hkg}. Because of the stability requirement, i.e. $F_a\gtrsim10^{17}\,{\rm GeV}$, there is no proper quintessence axion candidate. However, this warped geometry can provide us with an excellent fuzzy DM candidate. The fuzzy DM of mass $10^{-21}$--$10^{-19}\,{\rm eV}$~\cite{Irsic:2017yje,Armengaud:2017nkf,Ferreira:2020fam,Hui:2021tkt,Qiu:2022uvt} is very attractive, since we may naively understand the size of galaxies by its de Broglie wavelength. Furthermore, it may not have small-scale problems including the cusp-core problem. By adopting $f_1=f_2=M_*$, to get the correct fuzzy DM mass $m_a=10^{-19}\,{\rm eV}$, we can derive $F_a\simeq 8.3\times10^{15}\,{\rm GeV}$, which is consistent with the initial value of the fuzzy DM field to explain the DM density by its coherent oscillation.

We can build a QCD axion model in the present framework introducing an additional chiral $U(1)$ gauge symmetry where the additional fermions are two pairs of quarks $Q_i$ and antiquarks $\overline{Q}_i ~(i=1,2)$. The PQ breaking scale $F_a$ can be $10^{9-16}$ GeV. We see the required high quality to solved the strong $CP$ problem is maintained with $M_*\simeq 10^{16-17}$ GeV. It is remarkable that there is no serious domain wall problem even if the PQ symmetry breaking occurs after the inflation, since the domain wall number in the model is $N_{\rm DW}=1$.

The present model can also be easily embedded in a supersymmetric theory. In this case, we have dimension-5 proton decay operators \cite{Sakai:1981pk, Weinberg:1981wj} suppressed by only one power of the fundamental cutoff $M_* \simeq 10^{17}$ GeV. These enhanced proton decays will be tested in JUNO and Hyper-Kamiokande experiments\cite{Evans:2021hyx}.

\begin{acknowledgements}
T. T. Y. is supported in part by the China Grant for Talent Scientific Start-Up Project and by Natural Science Foundation of China (NSFC) under Grant No. 12175134 as well as by World Premier International Research Center Initiative (WPI Initiative), MEXT, Japan.
\end{acknowledgements}

\appendix
\section{Extra-dimensional suppression from bulk propagator}\label{app:A}
In this appendix, we provide details on the exponential suppression in the effective theory noted in the text as a result of integrating out the heavy mediator and the fifth dimension. 

First, we discuss the case for the flat extra dimension. Starting with the action in Eq.~(\ref{eq:action}), we obtain the classical equation of motion for $\Phi$, given in Eq.~(\ref{eq:EOM_Phi}). Utilizing this equation of motion, and plugging back into the action, we can integrate out the extra dimension $y$. This generates the effective four-dimensional interaction terms for $\phi_{1,2}$. The leading order suppression factor comes from the zero-mode for the bulk propagator, namely 
\begin{align}
    \nonumber
    {\cal L}_{ \rm eff, int} &= 2 c_1 c_2  M_*^3 \int  dy  \ \phi_1 \delta(y) \phi_2 
    \left[ \frac{1}{-\partial_y^2 + M_*^2} \delta (y-L) \right] \\
    &= 2 c_1 c_2  M_*^3 \int_{-\infty}^{\infty} \frac{dk}{2\pi} \frac{\phi_1 \phi_2}{k^2 + M_*^2} \int dy \delta(y) e^{ik(y-L)} \ ,
\end{align}
where we have used the representation of one delta function in terms of its exponential integration, and it is implied that Hermitian conjugation is added. Now, the integral over the extra dimension gives rise to a factor $e^{-ikL}$. Hence, we obtain
\begin{equation}
{\cal L}_{\rm eff, int} = 2 c_1 c_2  M_*^3  \phi_1 \phi_2 \int_{-\infty}^{\infty} \frac{dk}{2\pi} \frac{ e^{-i k L}}{k^2 + M_*^2}
\end{equation}

The final integration over $k$ picks up the residue at the pole $k=-i M_*$, and the resultant four-dimensional effective interaction term is given by
    \begin{equation}
        {\cal L}_{\rm eff, int} =  c_1 c_2  M_*^2 \phi_1 \phi_2 e^{-M_* L} + {\rm H.c.} \ ,
    \end{equation}
which is Eq.~(\ref{eq:Leff}) in the main text. Note that this can be inferred just by looking at the form of the bulk propagator of $\Phi$ in the position space in the extra dimension. 

Moving on to the case of warped extra dimension, the background metric now depends on the extra dimension in a nonfactorizable way, namely
\begin{equation}
    ds^2 = g_{ab} dx^a dx^b = (kz)^{-2} (\eta_{\mu \nu} dx^\mu dx^\nu-dz^2)  \ ,
\end{equation}
where $k$ is the AdS curvature scale, $a,b$ runs from $0$-$4$, and $\mu, \nu$ ranges from $0$-$3$, and $z \equiv e^{ky}/k$ is the conformal coordinate. This background solution is obtained for appropriate choices for the brane tensions and the bulk cosmological constant. As in the flat case, the origin for the suppression factor can be understood by looking at the limits of the bulk propagator. This was studied in great detail by Ref.~\cite{Fichet:2019hkg}, which also considered the effects of the dressed propagator for the timelike mediator momentum. We sketch the argument in this appendix.

In the AdS background, the propagator is obtained from 
    \begin{equation}
       \left[\frac{1}{\sqrt{|g|}}  \partial_a \left( \sqrt{|g|} g^{ab} \partial_b \right)  + M_*^2 \right] \Delta (x,x') = -i \frac{\delta^{(5)}(x-x')}{\sqrt{|g|}} \ .
    \end{equation}
Further, depending on the brane localized potentials for $\Phi$, the propagator needs to satisfy appropriate boundary conditions. Moving to the mixed coordinate space $(p,z)$, where $\eta^{\mu \nu} \partial_\mu \partial_\nu \Phi = - \eta^{\mu \nu} p_\mu p_\nu \Phi$, and $p \equiv \sqrt{p_\mu p^\mu}$, the solution for the propagator between $z$, and $z'$ in the asymptotic limit of $|p| > 1/{\rm min}(z,z')$, is suppressed as 
\begin{equation}
    \Delta_p(z,z') \propto e^{-|{\rm Im}p||z-z'|} \ ,
\end{equation}
where this suppression is analogous for the flat space for spacelike propagators, and for the timelike propagator this imaginary part is generated from 1PI dressed propagator. Similar conclusions hold for a vector propagator, which in the $A_5=0$ gauge takes the form~\cite{Goldberger:2002cz}
\begin{equation}
    \Delta_{p;\mu \nu}(z,z') \simeq \frac{k z}{p} \frac{K_1 (pz')}{K_0(pz)} \eta_{\mu \nu} + {\rm pure \ gauge} \ ,
\end{equation}
and asymptotically, for $|p|  {\rm min}(z,z') \gg 1$, the Bessel-$K$ function behaves as
\begin{equation}
    K_{\alpha} (\eta) \simeq \sqrt{\frac{\pi}{2 \eta}} e^{-\eta} + {\cal O} (\alpha^2) \ .
\end{equation}
For application to our case, $M_*$ is identified with the mass of the IR brane. With above asymptotic form for the propagator, the interaction Lagrangian becomes
\begin{equation}
    {\cal L}_{\rm eff, int} \sim c_1 c_2 M_*^2 \phi_1 \phi_2 e^{-M_{\rm Pl}/M_*} + {\rm H.c.} \ ,
\end{equation}
where the bulk propagator is assumed at $M_{\rm Pl}$, and $|z-z'| = |1/M_{\rm Pl}-1/M_*| \simeq 1/M_*$.

\bibliography{reference}
\bibliographystyle{apsrev4-1}
\end{document}